\documentclass[pdftex]{article}
\usepackage{icrctc07}

\title{Monte Carlo studies of the VERITAS array of Cherenkov telescopes}
\shorttitle{MC Studies of VERITAS}
\authors{Gernot Maier$^{1}$ for the VERITAS Collaboration$^{2}$ }
\shortauthors{G.Maier et al.}
\afiliations{$^{1}$ McGill University, 3600 University Street, H3A 2T8 Montreal, QC, Canada, \\
             $^{2}$ For full author list see G.Maier, ``VERITAS: Status and Latest Results", these proceedings}
\email{maierg@physics.mcgill.ca}

\abstract
{
VERITAS is a system of four imaging Cherenkov
telescopes located at the Fred Lawrence Whipple Observatory in southern Arizona.
We present here results of detailed Monte Carlo simulations of the array response
to extensive air showers. Cherenkov image and shower parameter distributions are
calculated and show good agreement with distributions obtained from observations
of background cosmic rays and high-energy $\gamma$-rays.
Cosmic-ray and $\gamma$-ray rates are accurately predicted by the simulations.
The energy threshold of the 3-telescope system is about 150 GeV after $\gamma$-hadron separation cuts;
the detection rate after $\gamma$-selection cuts for the Crab Nebula is 7.5 $\gamma$s/min.
The three-telescope system is able to detect a source with a flux equivalent to 10\% of 
the Crab Nebula flux in 1.2 h of observations (5 $\sigma$ detection).
}

\begin{document}
\maketitle

\vspace{-0.35cm}
\section{Introduction}
\vspace{-0.25cm}

The complete VERITAS array of four imaging Cherenkov telescope has been in operation
since early 2007 at the Fred Lawrence Whipple Observatory in southern Arizona.
Several TeV $\gamma$-ray sources have already been detected and extensive reports
about these observations can be found at this conference (see \cite{VE07} and references therein).
A detailed description of the first VERITAS telescope can be found in \cite{JH05}, 
and a full description of the VERITAS project in \cite{Weekes02}.
Results regarding a comparison of single telescope parameters from simulations and data
can be found in \cite{MAI05}. 
This paper concentrates on the performance of the array of telescopes.

\vspace{-0.40cm}
\section{Monte Carlo simulations}
\vspace{-0.25cm}

The VERITAS collaboration uses several different simulation chains in order to be 
independent of possible systematic effects arising from individual simulation packages.
The shower generators generally used are KASCADE \cite{Sembroski} and CORSIKA \cite{Heck}, 
while the response of the array of telescopes is simulated with three different packages 
(ChiLa, KASCADE, GrISUDet \cite{LeBohec}).
A description of the results from all of the simulation packages is beyond the scope of this paper, 
and so we concentrate on the CORSIKA/GrISUDet chain.

CORSIKA version 6.501 is used with the hadronic interaction models QGSJet, for primary energies above 
500 GeV, and FLUKA at lower energies.
Gamma-ray, hydrogen and helium nuclei induced air showers 
are simulated in an energy range from 50 GeV (30 GeV for hydrogen) to 30 TeV at different elevations.
Spectral indices are taken for cosmic rays from fits to balloon measurements \cite{BESS,RUNJOB}. For
$\gamma$-rays a distribution similar to the energy spectrum of the Crab Nebula has been simulated.
The shower cores are scattered randomly on a circular area with a radius of 600 m around the center of the
telescope array.
The isotropic distribution of the cosmic-ray incident angles is simulated by randomizing the shower directions
in a cone of radius 3.5$^{\circ}$ around the pointing direction of the telescopes.


Measurements of the atmospheric properties at the site of VERITAS 
are in progress \cite{MD07b}.
The calculations described here use the U.S. standard atmosphere, which does not always
reflect the properties of the atmosphere in southern Arizona.
The photoelectron rate per PMT is measured to be 100-200 MHz, corresponding to a night sky background rate of
\mbox{$2.8 \times 10^{12}$ photons $\cdot$ m$^{-2}$ s$^{-1}$ sr$^{-1}$}, which is used in the simulations.

The telescope and array simulations consist of three parts: the propagation of Cherenkov photons through
the optical system, the response of the camera and electronics, and the local and array trigger system.
The geometrical properties of the optical system are fully implemented;
misalignment of the mirrors and their surface roughness are taken into account.
The camera for each telescope consists of 499 pixels with 1 1/8'' phototubes and 
light cones.
The response of the PMTs to single photons has been measured;
the single photo-electron pulse has a rise time of 3.3 ns and a width of 6.5 ns \cite{JH05}.
In the simulations a signal in a PMT is created by summing single photo-electron pulses
with appropriate amplitude and time jitter applied.
Electronic noise and all efficiencies, including mirror reflectivities, geometrical, quantum,
and collection efficiencies, and losses due to signal transmission have been modeled.
The pulses are digitized with 2 ns sampling and with a trace length of 24 samples, reflecting the properties of the FADC system.
The trigger simulation utilizes a simplified model of the constant fraction discriminator, which is the
first stage of the VERITAS multi-level trigger, and a full realization of the pattern trigger,
requiring three adjacent pixels above threshold in a time window of 5 ns.
The currently used trigger threshold of 50 mV corresponds to about 4.8 photoelectrons.
The third level of the trigger system, the array trigger, activates the read out of the telescopes when
at least two telescopes pass the pattern trigger requirements in a time window of 100 ns.
The analysis steps, which include pedestal calculation, image cleaning, image parameterisation, source reconstruction, 
and calculation of mean scale variables for $\gamma$/hadron separation 
are exactly the same for simulated and real data and are described elsewhere \cite{MD07, KR06}.

\vspace{-0.40cm}
\section{Data}
\vspace{-0.25cm}

For the following comparison of data with Monte Carlo simulations, $\gamma$-ray candidates are extracted
from  five hours of observations of the Crab Nebula in January and February 2007.
The data were taken with an array of three telescopes (Telescopes 1,2, and 3) in wobble mode (0.5$^{\mathrm{o}}$
to 1.4$^{\mathrm{o}}$ offsets)
at a telescope elevation of about 70$^{\mathrm{o}}$ in good weather conditions.
$\gamma$-ray candidates were selected from the data using cuts on the shape of the shower images 
(so-called mean scaled width and length cuts, see \cite{MD07}) and on the shower direction 
($\Theta^2$, the squared difference between reconstructed shower direction and source position in the sky).
These cuts result in a 52 $\sigma$ detection of the Crab Nebula and provide  
$\sim1500$ $\gamma$-ray candidate events.

\vspace{-0.40cm}
\section{Comparison of Data with Monte Carlo simulation}
\vspace{-0.25cm}

\begin{figure}
\centering
%
\includegraphics[width=0.42\textwidth]{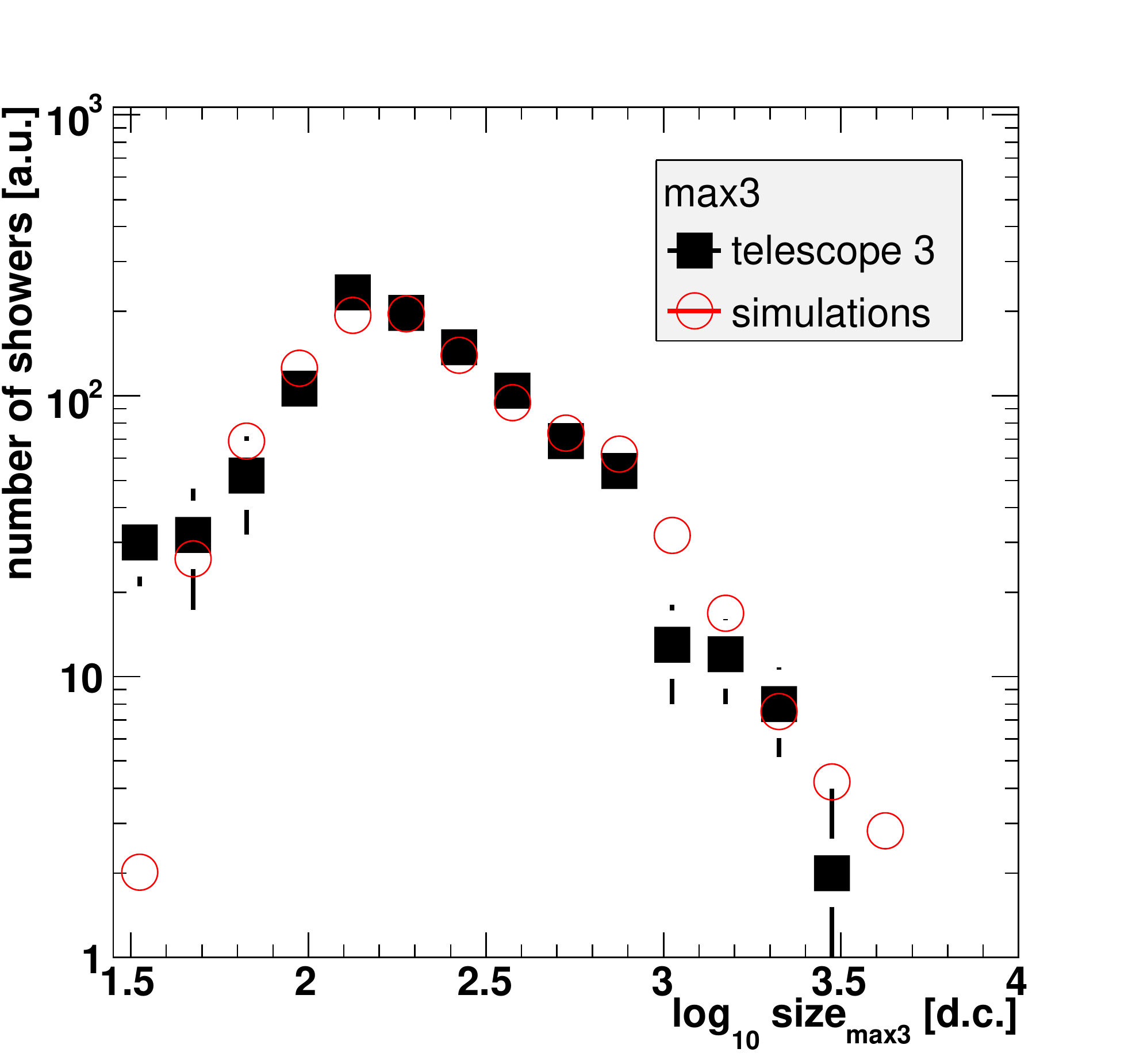}
\caption{\label{fig1} 
Comparison of integrated size in digital counts of the third highest pixel in images
in Telescope 3 for simulations (open symbols) and observation data (closed symbols).
}
\end{figure}

%
%
%
\begin{table}
\centering
\begin{tabular}{l|c|c|c}
   & wobble & data rate  & MC rate \\
   & offset & [$\gamma$/min]     & [$\gamma$/min] \\
\hline
2 tel & $0.3^{\mathrm{o}}$ & $4.4\pm0.1$ & $4.9\pm0.4$ \\
2 tel & $0.5^{\mathrm{o}}$ & $4.3\pm0.1$ & $4.7\pm0.4$ \\
3 tel & $0.5^{\mathrm{o}}$ & $7.5\pm0.1$ & $7.5\pm0.6$ \\
\end{tabular}
\caption{\label{tab:rates} Comparison of rates of $\gamma$-ray candidates
after cuts
for different telescope configurations and wobble offsets between data and simulations 
for observations of the Crab Nebula at elevations of about 70$^{\mathrm{o}}$.
The errors for the rate derived from real data is statistical only; the stated errors 
for the Monte Carlo rate reflect the uncertainty of the spectral parameters of the assumed
Crab Nebula energy spectrum.
}
\end{table}

\begin{figure}
\centering
%
%
%
%
%
\includegraphics[width=0.42\textwidth]{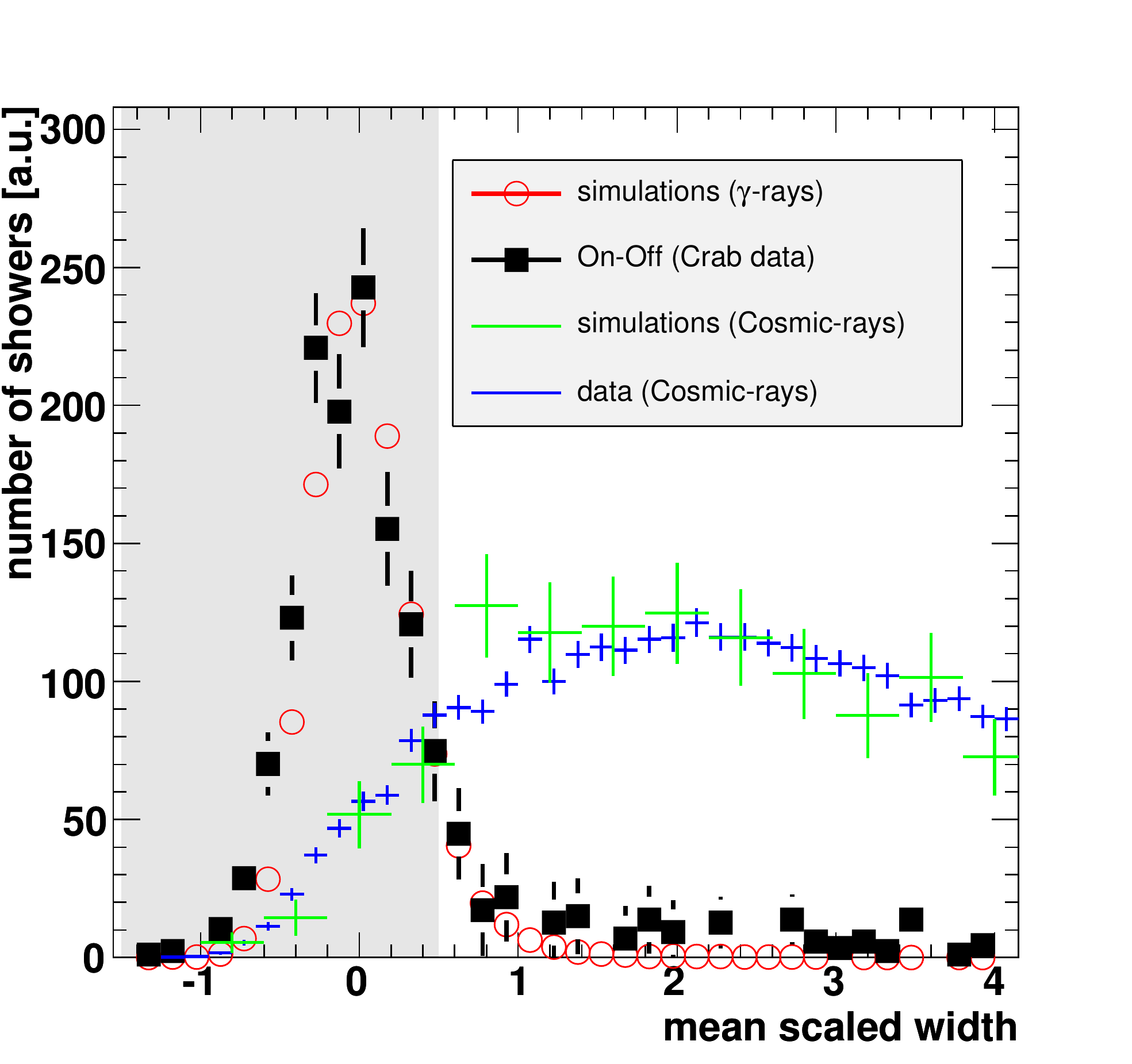}
\caption{\label{fig2} 
Comparison of mean scaled width for $\gamma$-ray and background data from simulations and 
observations.
The distributions have been scaled to similar fluxes.
The grey area indicates the range of values allowed by the $\gamma$-hadron separation cuts 
(same in Figure \ref{fig3}).
}
\end{figure}

\begin{figure}
\centering
\includegraphics[width=0.42\textwidth]{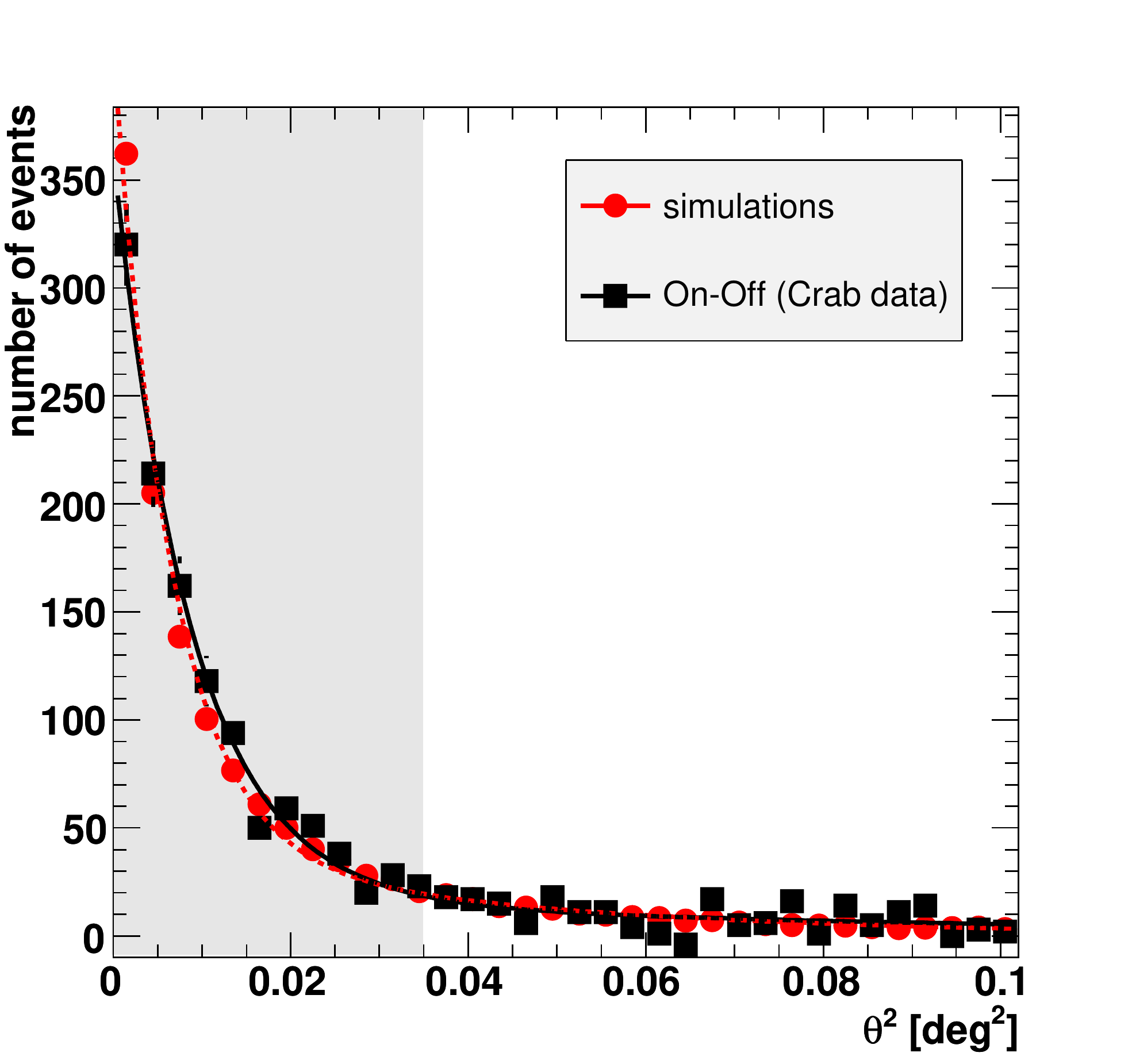}
\caption{\label{fig3} 
Comparison of $\Theta^{2}$ distribution from simulations (open symbols) and
real data (closed symbols).
The lines show fits to each of the distributions (dashed line for simulations, continuous line for data).
The distributions have been scaled to similar fluxes.
}
\end{figure}

\begin{figure}
\centering
\includegraphics[width=0.42\textwidth]{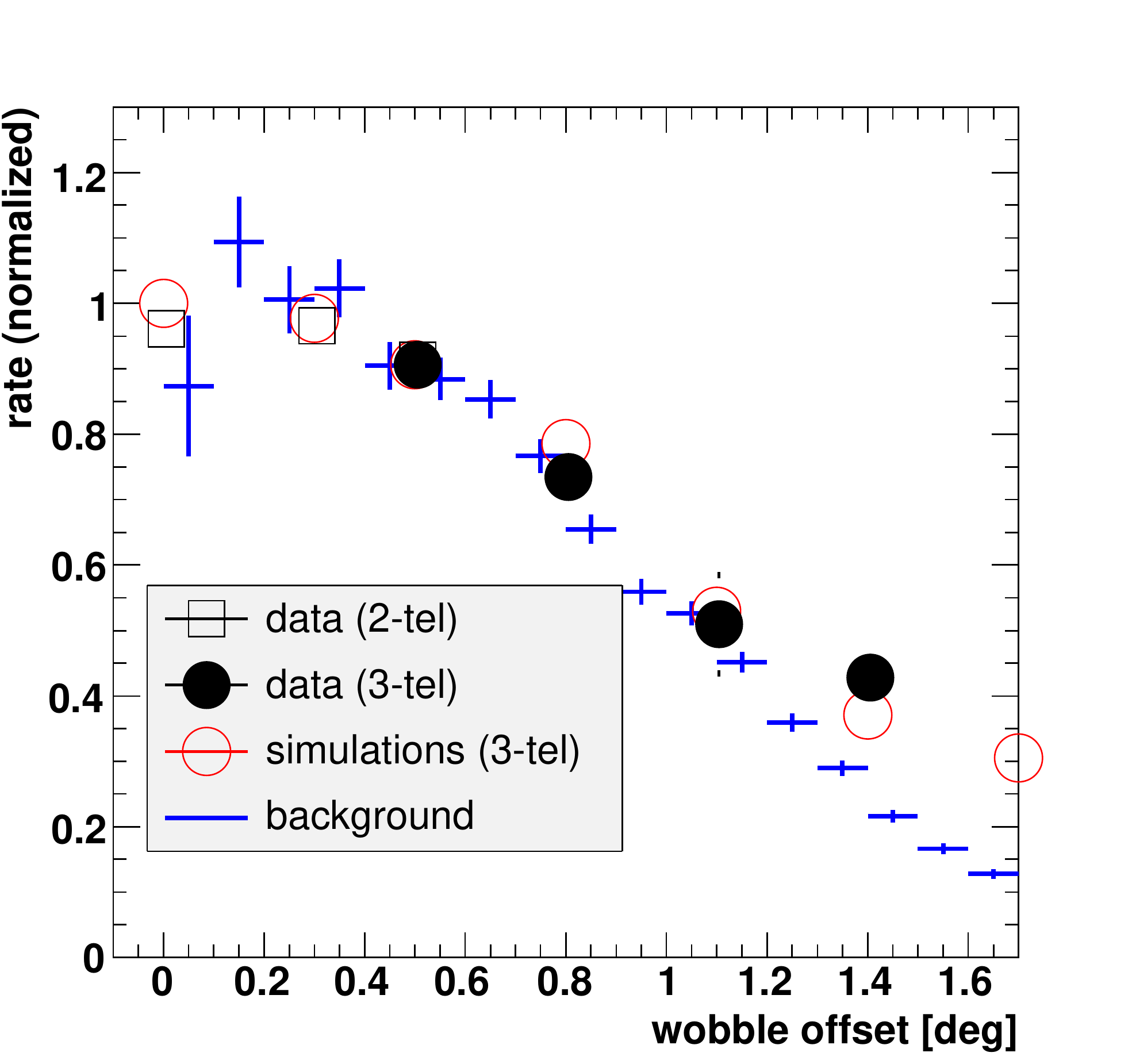}
\caption{\label{fig4} 
Relative $\gamma$-ray rate as function of distance to camera center for data (black symbols)
and simulations (red symbols) and $\gamma$-ray like showers (blue crosses).
}
\end{figure}

The pattern trigger 
of VERITAS requires at least three adjacent pixels above threshold in a certain time window.
The pixel with the third highest signal size in a pattern therefore determines the activation
of the readout of the telescope.
The comparison between data and simulations of the distribution
of the integrated charge measured in this pixel over many showers (Figure \ref{fig1}) demonstrates
that the second level trigger simulation reflects the real system accurately.
The event rate after basic quality cuts (requiring the successful reconstruction of shower direction
and impact parameter) is 22.3$\pm$0.3 Hz for data and 20.6$\pm$0.4 Hz for simulated cosmic-ray showers.
The typical dead time of the 3-telescope system of 6-7\% is taken into account.
Before comparing these numbers for $\gamma$-rays,
the parameters used for the suppression of background events have to be tested.
Figure \ref{fig2} shows the mean scaled width distributions for measured and simulated
$\gamma$-ray and
background events, Figure \ref{fig3} the $\Theta^2$ distributions for 
data and simulations.
The good agreement in both figures indicates that the simulations reflect the real system well.
The rate of detected $\gamma$-rays from the Crab Nebula depends on the applied
cuts. Those applied here are relatively hard: about 50\% of all $\gamma$-rays are accepted 
and more than 99.9\% of all cosmic-rays are rejected.
Table \ref{tab:rates} lists
$\gamma$-ray rates for different array configurations and wobble offsets for measurements 
and simulations.
The off-axis acceptance of the system is particularly important for the detection of extended
sources or during sky surveys. 
Figure \ref{fig4} shows that the acceptance is above 80\%, relative to a source at the center of the camera, for offsets from the camera center of up to 
0.7$^{\mathrm{o}}$ and is well described by the simulations.

%
%

%

%
%
%
\begin{figure}
\centering
\includegraphics[width=0.42\textwidth]{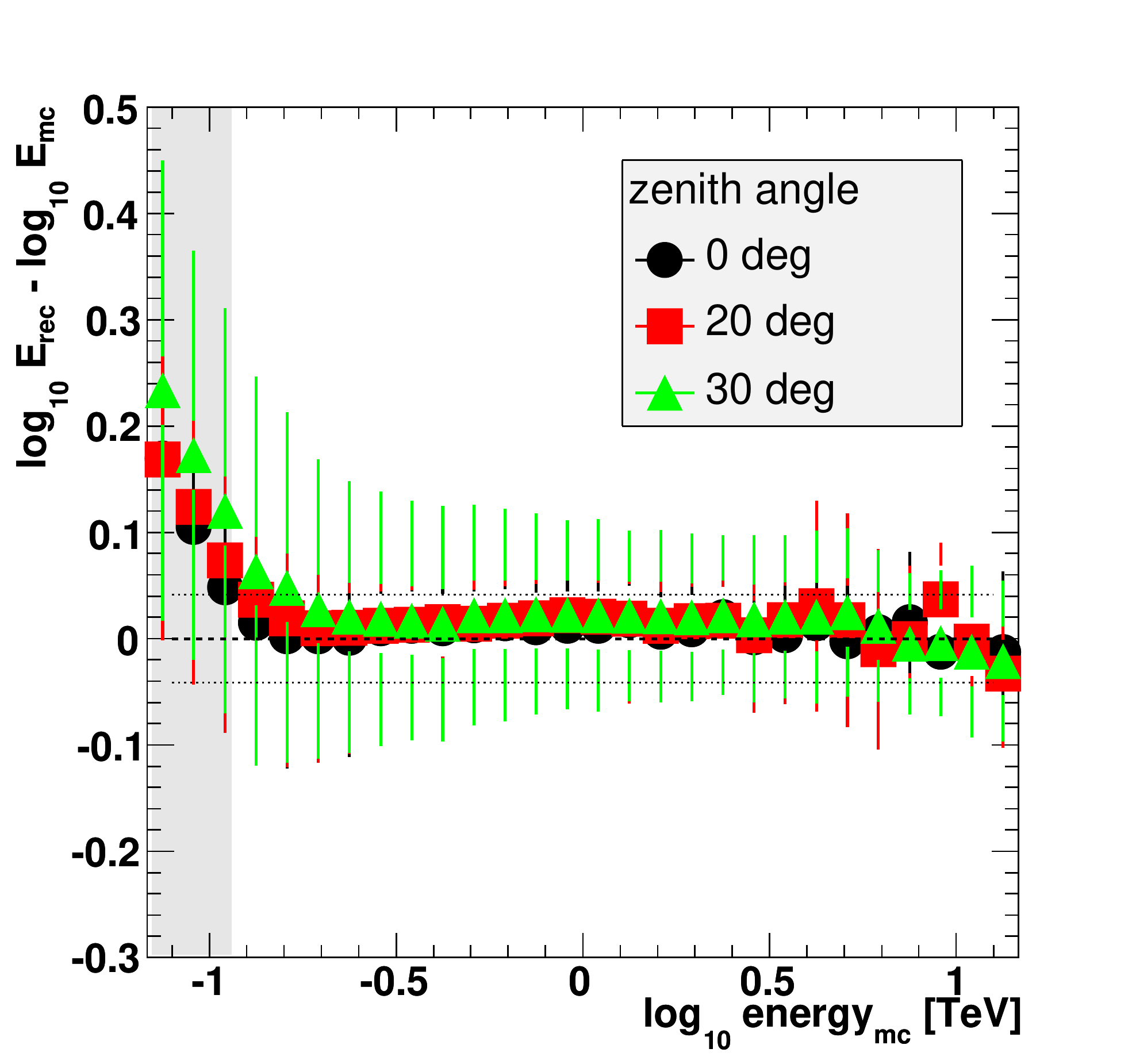}
\caption{\label{fig5}
Bias in energy reconstruction vs true energy E$_{MC}$
for three different elevations.
The error bars show the width of corresponding bias distribution
(for the array configuration T1+T2+T3 and an observation mode of  
wobble offset $0.5^{\mathrm{o}}$).
The dotted lines indicate a bias of 10\%.
}
\end{figure}

\begin{figure}
\centering
\includegraphics[width=0.42\textwidth]{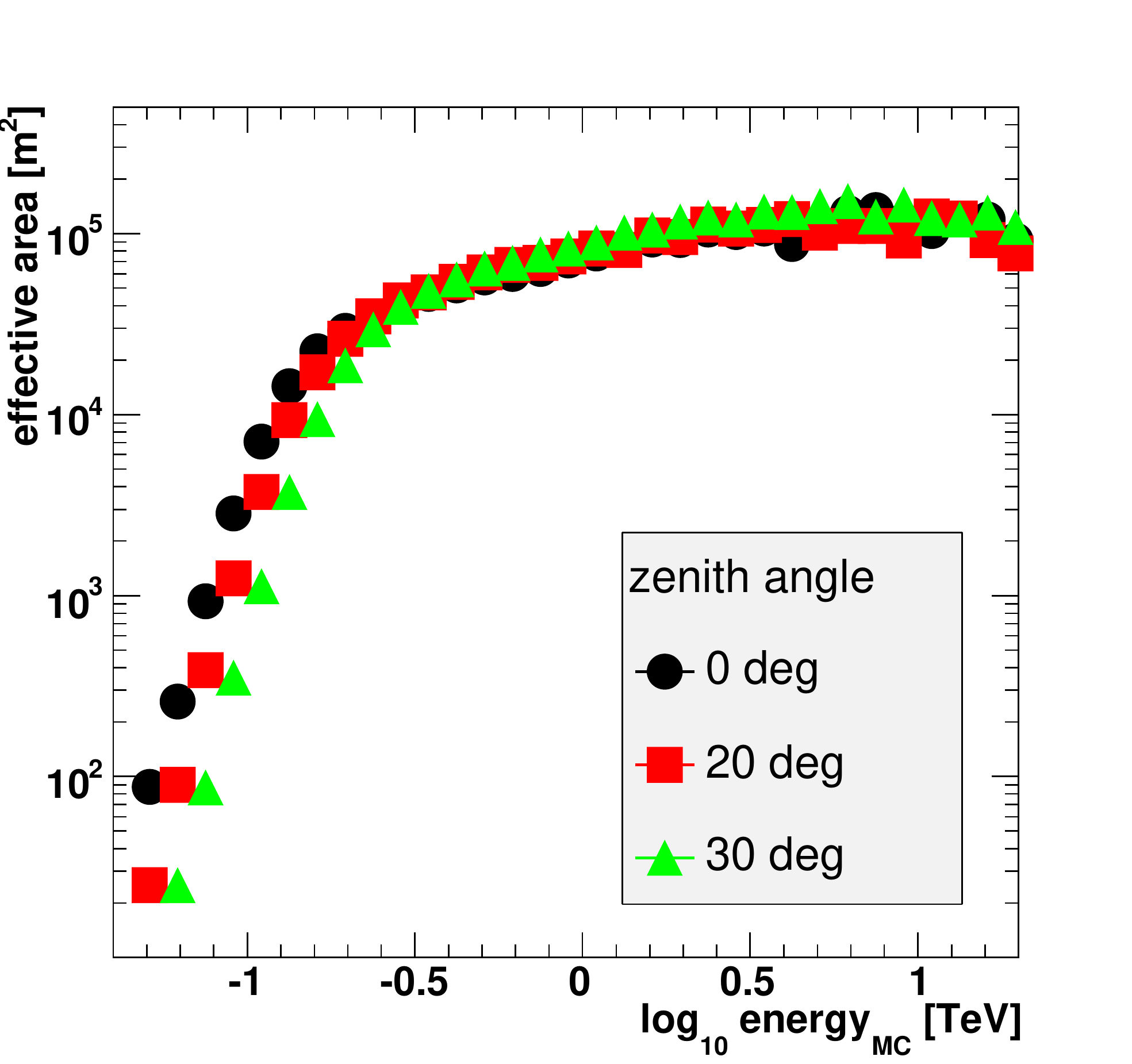}
\caption{\label{fig6}
Effective area vs true energy E$_{MC}$ for different elevations
(for the array configuration T1+T2+T3 and an observation mode of 
wobble offset $0.5^{\mathrm{o}}$).
}
\end{figure}

The energies of the primary $\gamma$-rays are reconstructed using lookup tables \cite{KR06}.
The bias in the reconstructed energy is, as Figure \ref{fig5} shows, well below 10\%
for energies above 160 GeV.
The effective area of the system is uniform above these energies (Figure \ref{fig6})
and extends well beyond 10 TeV.
The current analysis energy threshold of the system, defined as the position of the peak of the energy spectrum
of the source convolved with the effective area curve of the detector,
is about 150 GeV for observations near the zenith.
It increases to 300 GeV at 40$^{\mathrm{o}}$ zenith angle.
The three-telescope system is able to detect a source with a flux equivalent to 10\% of the flux of
the Crab Nebula in 1.2 h of observations (5 $\sigma$ detection).

\vspace{-0.45cm}
\section{Conclusions}
\vspace{-0.25cm}

Monte Carlo simulations of the VERITAS array of Cherenkov telescopes
show that the system is well understood and accurately modeled.
The agreement between the predicted and actual performance of the system, as well as the
detection of several sources of high-energy $\gamma$-rays during the construction phase of the project,
demonstrates the high potential of VERITAS for making detailed morphological and spectral studies 
of astrophysical $\gamma$-ray sources from the northern hemisphere.

\vspace{-0.45cm}
\subsection*{Acknowledgments}
\vspace{-0.25cm}

This research is supported by grants from the U.S. Department of Energy,
the U.S. National Science Foundation,
and the Smithsonian Institution, by NSERC in Canada, by PPARC in the UK and
by Science Foundation Ireland.


\vspace{-0.45cm}
\begin{thebibliography}{99}
\begin{small}
\bibitem{VE07} Maier, G. et al, 2007, {\em VERITAS: Status and Latest Results}, 30th ICRC, Merida
\bibitem{JH05} Holder J. et al., 2006, Astrop.Phys., 25, 391
\bibitem{Weekes02} Weekes, T. et al., 2002, Astroparticle Phys., 17, 221 
\bibitem{MAI05} Maier G. et al., 2005, {\em Monte Carlo Studies of the first VERITAS telescope }, 29th ICRC, Pune 
\bibitem{Sembroski} Kertzman M.P., Sembroski G.H., 1994, NIM A, 343, 629 
\bibitem{Heck} Heck D. et al., 1998, Report FZKA 6019, Forschungszentrum Karls\-ruhe
\bibitem{LeBohec} \mbox{Duke C., LeBohec S.}, http://www.physics.utah.edu/gammaray/GrISU/
\bibitem{MD07b}  Daniel M. et al., 2007, {\em Application of radiosonde data to VERITAS simulations},  30th ICRC, Merida
\bibitem{BESS} Haino S. et al., 2004, Phys.Letters B 594, 35
\bibitem{RUNJOB} Derbina V.A. et al., 2005, ApJ, 628 L41
\bibitem{MD07} Daniel M. et al., 2007, {\em The VERITAS Standard Data Analysis}, 30th ICRC, Merida
\bibitem{KR06} Krawczynski H. et al, 2006, Astroparticle Phys., 25, 380
\end{small}
\end{thebibliography}
\end{document}